\documentclass[aps,prd,11pt,superscriptaddress,amssymb,amsmath,nofootinbib]{revtex4-1}

\usepackage{color}
\usepackage{epsfig}
\usepackage{graphicx}
\usepackage{float}
\usepackage{epstopdf}
\usepackage{hyperref}
\usepackage[usenames,dvipsnames]{xcolor}
\usepackage{multirow}
\usepackage[normalem]{ulem}
\usepackage[d]{esvect}

\begin{document}

 \title{The Nonsymmetric Flavor Transition Matrix and the Apparent P violation}
 \author{Shu-Jun Rong}\email{rongshj@glut.edu.cn}
 \author{Ding-Hui Xu}\email{3036602895@qq.com}
 \affiliation{College of Science, Guilin University of Technology, Guilin, Guangxi 541004, China}

 \begin{abstract}
The leptonic mixing parameters of high precision and the next-generation neutrino telescopes
make it possible to test  new physics in the flavor transition of the high-energy astrophysical neutrinos(HAN).
We introduce a nonsymmetric matrix to modify the predictions of the standard flavor transition matrix.
It is constructed with the mixing matrix in vacuum and that at the source of the HAN. The mismatch of the mixing matrices
results in the new expectation of the flavor ratio of the HAN at Earth. It also leads to a
secondary effect called the apparent P violation(APV). The quantitative analyses of the
new effects are performed with a moderate setup of the parameters at the source of the HAN.
The correlations between the mixing parameters and the new predictions are shown.
From the correlations, the dominant parameters determining the new-physics effects are identified.

\end{abstract}


 \maketitle
 \section{Introduction}

Since the late 1990s, notable progresses have been made in the measurement of neutrino oscillation parameters\cite{1,2,3}.  According to the global analysis NuFIT 5.0  with the Super-Kamiokande atmospheric neutrino data, the $3\sigma$ relative precision of leptonic mixing angles and Dirac CP phase  is shown as follows\cite{4} :
\begin{equation}
\label{eq:1}
\theta_{12}:14\%,~~\theta_{13}:9\%,~~ \theta_{23}:25\%,~~\delta_{CP}:100\%.
\end{equation}
Obviously, we have the well measured $\theta_{13}$ and the uncertain $\delta_{CP}$. In the following decades, the precision of the mixing parameters will be improved by results from the oscillation experiments such as JUNO\cite{5}, Hyper-Kamiokande(HK)\cite{6}, DUNE\cite{7}, etc. Then new physics on neutrinos may be tested.
As we know, new physics usually lies in the extreme environment or the extreme parameter-range.
Recently, a large number of high-energy astrophysical neutrinos (HAN) events in the energy range of Tev-Pev have been reported by the IceCube Collaboration \cite{8,9,10,11}.
These events may provide information on new physics.

In this paper, we are concerned about the effects of new physics in the flavor transition of the HAN.
Because the HAN detected at Earth have traveled cosmological-scale distances, the decoherence effects of flavor neutrinos should be considered.
So the flavor conversion probability $\overline{P}_{\alpha\beta}$ is of the form
 \begin{equation}
\label{eq:2}
\overline{P}_{\alpha\beta}=|U_{\alpha1}|^{2}|U_{\beta1}|^{2}+|U_{\alpha2}|^{2}|U_{\beta2}|^{2}+|U_{\alpha3}|^{2}|U_{\beta3}|^{2},
\end{equation}
where $\alpha,~\beta=e,~\mu,~\tau$. $U_{\alpha i}(i=1,2,3)$ is the element of the leptonic mixing matrix\cite{12}. The so-called standard flavor transition matrix $\overline{P}$ is symmetric,
which is obtained on the base of the assumption that the mixing matrix of the HAN at their source and that in their journey are the same as that in vacuum, namely $U$. Any scenario beyond the assumption expressed by a different
flavor transition matrix could manifest a class of new physics. Typical new physics  includes the energy-dependent leptonic mixing matrix\cite{13,14}, neutrino-decay\cite{15,16,17,18,19,20,21,22}, non-unitary mixing\cite{23,24,25,26,27,28,29,30}, pseudo-Dirac neutrinos\cite{31,32,33,34,35,36}, neutrino secret interactions\cite{37,38,39,40}, etc\cite{41}.
These new-physics based transition matrices predict various of flavor ratios of the HAN at Earth. However, correlations between the parameters in the transition matrix and the new predictions  are usually unclear in these scenarios. In this paper, we propose a simplified nonsymmetric flavor transition matrix and analyse the relationship between the new effects and the mixing parameters.

In principle, matter interactions\cite{42,43,44,45}, nonstandard interactions\cite{46,47,48}, and other new physics may modify the leptonic mixing matrix at the source of the HAN.
Similar to the case of the solar neutrinos\cite{49}, we suppose that the HAN in their motherland undergo an adiabatic flavor evolution. Then we can generalize the flavor transition probability as
\begin{equation}
\label{eq:3}
\overline{P}^{g}_{\alpha\beta}=|U_{\alpha1}^{S}|^{2}|U_{\beta1}|^{2}+|U_{\alpha2}^{S}|^{2}|U_{\beta2}|^{2}+|U_{\alpha3}^{S}|^{2}|U_{\beta3}|^{2}.
\end{equation}
 Here $U^{S}$ is the leptonic mixing matrix at the location where the HAN are produced.
Although $\overline{P}^{g}$ is not of the most general form,
it could describe several classes of nonstandard flavor conversion scenarios in the adiabatic cases.
As an example, when $U^{S}$ is unitary, it could denotes a mixing matrix including the effect of matter interaction or
nonstandard interactions. In this paper, we focus on the model-independent properties and predictions of $\overline{P}^{g}$.

 In general, the flavor transition matrix $\overline{P}^{g}$ is nonsymmetric. It may give different predictions on
 the HAN flavor ratio at Earth. Furthermore, suppose that we interchange the departure location and the arrival location of neutrinos. Namely, we make a P transformation
 $|U_{\alpha i}^{S}|^{2}|U_{\beta i}|^{2}\longrightarrow |U_{\alpha i}|^{2}|U_{\beta i}^{S}|^{2}$, with i=1, 2, 3. The transition probability of the $\alpha$-flavor neutrinos departing from  Earth would
 be expressed as $(\overline{P}^{g})^{P}_{\alpha\beta}=\overline{P}^{g}_{\beta\alpha}$. In the case $\alpha\neq \beta$, we have $\overline{P}^{g}_{\alpha\beta}\neq \overline{P}^{g}_{\beta\alpha}$. So an apparent P violation (APV) quantified by $\overline{P}^{g}_{\alpha\beta}-\overline{P}^{g}_{\beta\alpha}$ would be observed. The APV means that the P violation is from the environment where neutrinos are produced. Furthermore, we note that
 $(\overline{P}^{g})^{T}_{\alpha\beta}=\overline{P}^{g}_{\beta\alpha}$. So the APV is equal to the apparent T violation (ATV).

In practice, the new effects from $\overline{P}^{g}$ are impacted by the uncertainties of the parameters in $U^{S}$. We have no precision information on  $U^{S}$
which is dependent on the specific source of the HAN. So we employ a model-independent method to
describe $U^{S}$.  In the standard parametrization\cite{12}, $|U^{S}|$ is dependent on the mixing angles $\theta_{ij}^{S}$($ij=12,23,13$) and the CP phase $\delta^{S}$.
We consider the case that the relative deviations of these parameters from their counterparts in vacuum are moderate, i.e., of the order of $10\%$. These setups
of $U^{s}$ enable us to find out the correlation between the new predictions and the leptonic mixing parameters.
These observations may hold for neutrinos produced in the realistic astrophysical sources such as galactic halos and cluster halos where the matter effect
 on $U^{S}$ is modest\cite{55}.

The paper is organised as follows. In Sec.~II, the properties of the generalized flavor transition matrix $\overline{P}^{g}$ are studied.
 On the basis of these properties, a generalized area theorem of the flavor triangle at Earth is derived. It reveals how the flavor ratio distribution at Earth is dependent on the determinant of $\overline{P}^{g}$.
 The expression of the APV is given in this section. The possible case that the APV may be detected is discussed.
 In Sec.~III, we perform numerical analyses of the impacts of $\overline{P}^{g}$. The correlation  between the determinant of the matrices $\overline{P}^{g}$ and that of $\overline{P}$ is given.
 The quantitative flavor ratio distribution at Earth on the basis of  $\overline{P}^{g}$ and the special initial flavor ratios are obtained.
 The correlations between the APV and the mixing parameters are analysed. Finally, we conclude.

\section{The generalized flavor transition matrix and new-physics effects}

\subsection{Flavor ratio vector and flavor triangle}
\begin{figure}[t]
\centering 
\includegraphics[width=.49\textwidth]{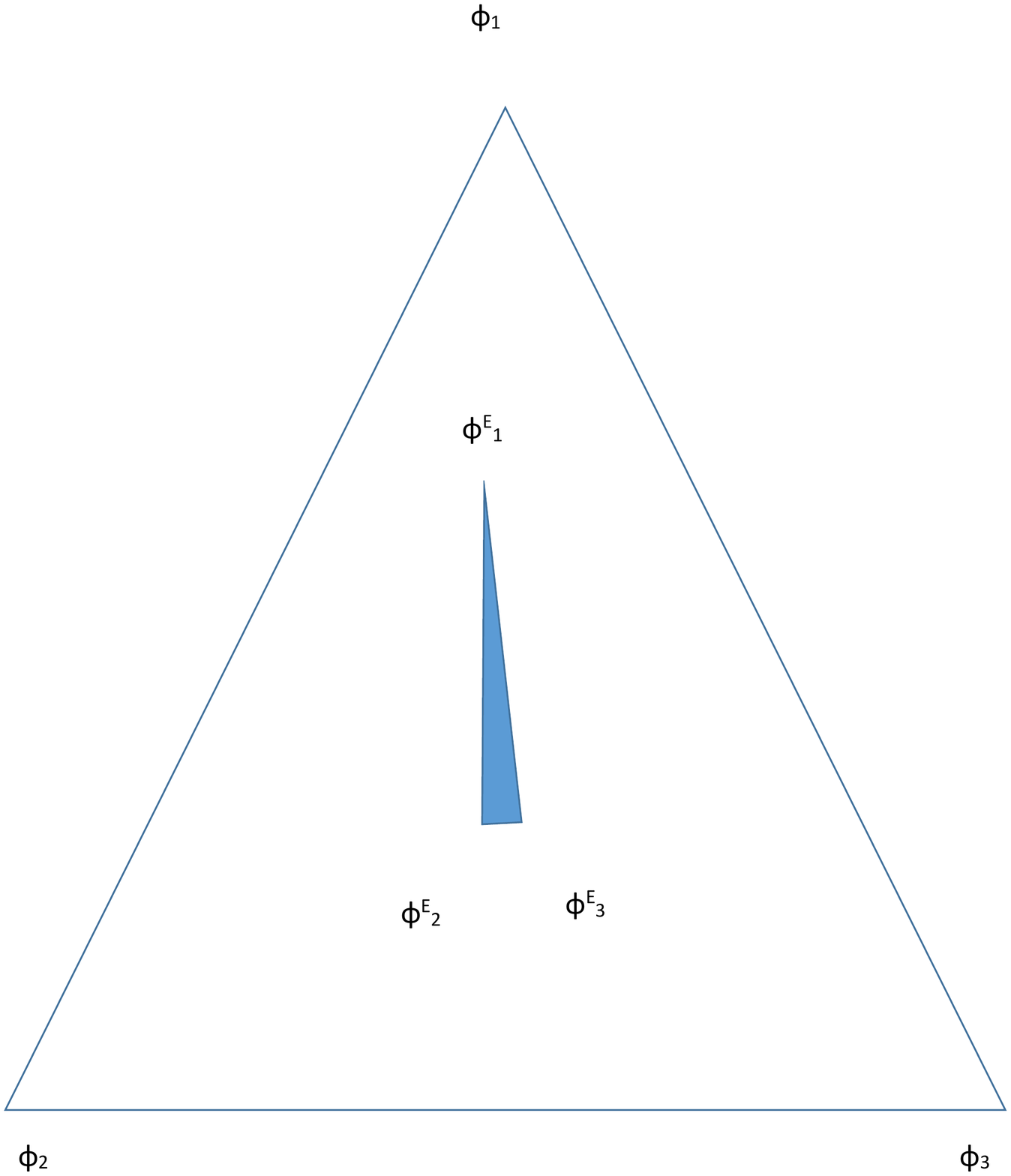}
\caption{\label{fig:1} The schematic figure of the flavor triangle and the Earth flavor triangle (EFT).  }
\end{figure}

The general neutrino flavor ratio could be expressed with a column vector, namely
\begin{equation}
\label{eq:4}
\overrightarrow{\phi}=\left(
                                                     \begin{array}{ccc}
                                                      \phi_{e} & \phi_{\mu} & \phi_{\tau} \\
                                                     \end{array}
                                                   \right)^{T},
\end{equation}
which satisfies the following conditions
\begin{equation}
\label{eq:5}
\phi_{e} + \phi_{\mu} + \phi_{\tau}=1,~~0\leq\phi_{\alpha}\leq1.
\end{equation}
So the set of the points of all the flavor ratio vectors forms a triangle in the 3-dimensional Euclidean space.  Follows the convention of the Ref.\cite{50}, we call it flavor triangle, see Fig.\ref{fig:1}.
The vertexes of the flavor triangle correspond to the vectors of the form
\begin{equation}
\label{eq:6}
\overrightarrow{\phi_{1}}=\left(
                                                     \begin{array}{ccc}
                                                     1 & 0 & 0 \\
                                                     \end{array}
                                                   \right)^{T},~~ \overrightarrow{\phi_{2}}=\left(
                                                     \begin{array}{ccc}
                                                    0 &1 & 0 \\
                                                     \end{array}
                                                   \right)^{T}, ~~ \overrightarrow{\phi_{3}}=\left(
                                                     \begin{array}{ccc}
                                                    0 &0 & 1 \\
                                                     \end{array}
                                                   \right)^{T}.
\end{equation}
The area of the flavor triangle is $S_{\triangle_{FT}}=\sqrt{3}/2$.
Note that $\overrightarrow{\phi_{1}}$ ,  $\overrightarrow{\phi_{2}}$ correspond to the flavor ratio at the neutron-decay source and that at the muon-damping source respectively.

Suppose that the flavor ratio of the HAN at the source is given by $\overrightarrow{\phi^{S}}$. When the HAN arrive at Earth, $\overrightarrow{\phi^{S}}$ is changed to  $\overrightarrow{\phi^{E}}$.
It is connected to $\overrightarrow{\phi^{S}}$ through the flavor transition matrix, namely
\begin{equation}
\label{eq:7}
\left(
  \begin{array}{c}
    \phi_{e}^{E} \\
   \phi_{\mu}^{E} \\
    \phi_{\tau}^{E}\\
  \end{array}
\right)
=\left(
    \begin{array}{ccc}
    \overline{P}_{ee}^{g} & \overline{P}_{e\mu}^{g} &  \overline{P}_{e\tau}^{g} \\
      \overline{P}_{\mu e}^{g} &  \overline{P}_{\mu\mu}^{g} & \overline{P}_{\mu\tau}^{g} \\
     \overline{P}_{\tau e}^{g} &  \overline{P}_{\tau\mu}^{g} & \overline{P}_{\tau\tau}^{g}, \\
    \end{array}
  \right)\left(
  \begin{array}{c}
    \phi_{e}^{S} \\
   \phi_{\mu}^{S} \\
    \phi_{\tau}^{S}\\
  \end{array}
\right).
\end{equation}
Here the expression of $\overline{P}_{\alpha\beta}^{g}$ is shown in Eq.\ref{eq:3}. After the flavor transition of HAN, the flavor triangle is transformed to that at Earth.
We call it  Earth flavor triangle (EFT). The schematic figure of the EFT is also shown in Fig.\ref{fig:1}.
Given the matrix $\overline{P}^{g}$, the EFT shows the general flavor ratio distribution of the HAN at Earth.
The vertexes of the EFT correspond to the vectors $\overrightarrow{\phi^{E}_{1}}$, $\overrightarrow{\phi^{E}_{2}}$, and $\overrightarrow{\phi^{E}_{3}}$.
They are the first, the second and the third column vector of the flavor transition matrix $\overline{P}^{g}$ respectively, namely
\begin{equation}
\label{eq:8}
\overrightarrow{\phi^{E}_{1}}=\left(
                                \begin{array}{c}
                                   \overline{P}_{ee}^{g} \\
                                   \overline{P}_{\mu e}^{g}\\
                                 \overline{P}_{\tau e}^{g}  \\
                                \end{array}
                              \right),~~\overrightarrow{\phi^{E}_{2}}=\left(
                                \begin{array}{c}
                                   \overline{P}_{e\mu}^{g} \\
                                  \overline{P}_{\mu \mu}^{g}\\
                                 \overline{P}_{\tau \mu}^{g}  \\
                                \end{array}
                              \right),~~\overrightarrow{\phi^{E}_{3}}=\left(
                                \begin{array}{c}
                                  \overline{P}_{e\tau}^{g} \\
                                   \overline{P}_{\mu \tau}^{g}\\
                                 \overline{P}_{\tau \tau}^{g}  \\
                                \end{array}
                              \right).
\end{equation}
The area of the EFT is determined by the determinant of $\overline{P}^{g}$. The specific expression is shown in the following part.

\subsection{Unitarity and the  generalized area theorem for the EFT}

In this paper, we consider the case that the mixing matrices $U$ and $U^{S}$ are unitary. So the columns vectors
$\overrightarrow{\phi^{E}_{1}}$, $\overrightarrow{\phi^{E}_{2}}$, $\overrightarrow{\phi^{E}_{3}}$ satisfy the constraints
\begin{equation}
\label{eq:9}
\sum_{i}\phi^{E}_{1i}=\sum_{i}\phi^{E}_{2i}=\sum_{i}\phi^{E}_{3i}=1,
\end{equation}
\begin{equation}
\label{eq:10}
\overrightarrow{\phi^{E}_{1}}+\overrightarrow{\phi^{E}_{2}}+\overrightarrow{\phi^{E}_{3}}=3\overrightarrow{o},
\end{equation}
where $\overrightarrow{o}$ is the democratic vector of the form
\begin{equation}
\label{eq:11}
\overrightarrow{o}=\left(
                     \begin{array}{ccc}
                       1/3 & 1/3 &1/3 \\
                     \end{array}
                   \right)^{T}.
\end{equation}
On the basis of unitarity conditions Eq.\ref{eq:9} and Eq.\ref{eq:10},
we can derive an identity for the column vectors of $\overline{P}^{g}$, namely
\begin{equation}
\label{eq:12}
\overrightarrow{\phi^{E}_{1}}\times\overrightarrow{\phi^{E}_{2}}+\overrightarrow{\phi^{E}_{2}}\times\overrightarrow{\phi^{E}_{3}}
+\overrightarrow{\phi^{E}_{3}}\times\overrightarrow{\phi^{E}_{1}}=
Det(\overline{P}^{g})(\overrightarrow{\phi^{E}_{1}}+\overrightarrow{\phi^{E}_{2}}+\overrightarrow{\phi^{E}_{3}})=3Det(\overline{P}^{g})\overrightarrow{o}.
\end{equation}
Here Det denotes the determinant, $Det(\overline{P}^{g})$=$(\overrightarrow{\phi^{E}_{1}}\times\overrightarrow{\phi^{E}_{2}})\cdot\overrightarrow{\phi^{E}_{3}}$.
The dot product and cross product of flavor vectors are defined in the standard way, namely
\begin{equation}
\label{eq:13}
\overrightarrow{a}\cdot\overrightarrow{b}=\sum_{i}a_{i}b_{i}, ~~\overrightarrow{d}=\overrightarrow{a}\times\overrightarrow{b},
~~with ~~d_{i}=\sum_{j, k}\epsilon_{ijk}a_{j}b_{k},
\end{equation}
where $\epsilon_{ijk}$ is the totally antisymmetric tensor  with $\epsilon_{123}=1$.
The proof of the identity is given as follows.

In the case that the column vectors are linearly independent, namely $Det(\overline{P}^{g})\neq0$,
we can  decompose the circular cross product of the vectors as
\begin{equation}
\label{eq:14}
\overrightarrow{\phi^{E}_{1}}\times\overrightarrow{\phi^{E}_{2}}+\overrightarrow{\phi^{E}_{2}}\times\overrightarrow{\phi^{E}_{3}}+\overrightarrow{\phi^{E}_{3}}\times\overrightarrow{\phi^{E}_{1}}=
x\overrightarrow{\phi^{E}_{1}}+y\overrightarrow{\phi^{E}_{2}}+z\overrightarrow{\phi^{E}_{3}}.
\end{equation}
Then employing the identity
\begin{equation}
\label{eq:15}
(\overrightarrow{\phi^{E}_{1}}\times\overrightarrow{\phi^{E}_{2}})\cdot\overrightarrow{\phi^{E}_{3}}=
(\overrightarrow{\phi^{E}_{2}}\times\overrightarrow{\phi^{E}_{3}})\cdot\overrightarrow{\phi^{E}_{1}}=
(\overrightarrow{\phi^{E}_{3}}\times\overrightarrow{\phi^{E}_{1}})\cdot\overrightarrow{\phi^{E}_{2}},
\end{equation}
we obtain the following equations
\begin{equation}
\label{eq:16}
(\overline{P}^{g})^{T}\overline{P}^{g}\left(
         \begin{array}{c}
           x \\
           y \\
           z \\
         \end{array}
       \right)=
\left(
  \begin{array}{ccc}
    \overrightarrow{\phi^{E}_{1}}\cdot\overrightarrow{\phi^{E}_{1}} &\overrightarrow{\phi^{E}_{1}}\cdot\overrightarrow{\phi^{E}_{2}} &\overrightarrow{\phi^{E}_{1}}\cdot\overrightarrow{\phi^{E}_{3}} \\
  \overrightarrow{\phi^{E}_{2}}\cdot\overrightarrow{\phi^{E}_{3}} &\overrightarrow{\phi^{E}_{2}}\cdot\overrightarrow{\phi^{E}_{2}} & \overrightarrow{\phi^{E}_{2}}\cdot\overrightarrow{\phi^{E}_{3}}\\
    \overrightarrow{\phi^{E}_{3}}\cdot\overrightarrow{\phi^{E}_{1}} &\overrightarrow{\phi^{E}_{3}}\cdot\overrightarrow{\phi^{E}_{2}} & \overrightarrow{\phi^{E}_{3}}\cdot\overrightarrow{\phi^{E}_{3}}\ \\
  \end{array}
\right)\left(
         \begin{array}{c}
           x \\
           y \\
           z \\
         \end{array}
       \right)=3Det(\overline{P}^{g})\overrightarrow{o}.
\end{equation}
According to the constraints Eq.\ref{eq:9} and  Eq.\ref{eq:10}, we find that $\overrightarrow{o}$
is  the eigenvector of the coefficient matrix $(\overline{P}^{g})^{T}\overline{P}^{g}$ with the eigenvalue equal to 1. Because of the condition $Det(\overline{P}^{g})\neq0$ , we obtain the result $x=y=z=Det(\overline{P}^{g})$.
So the identity is proved.

 Now we consider the case that column vectors are linearly dependent, namely $Det(\overline{P}^{g})=0$. On the basis of the constraints  Eq.\ref{eq:9} and  Eq.\ref{eq:10},
 vectors $\overrightarrow{\phi^{E}_{2}}$ and $\overrightarrow{\phi^{E}_{3}}$ could be decomposed as
\begin{eqnarray}
\label{eq:17}
 \overrightarrow{\phi^{E}_{2}}&=& x_{1} \overrightarrow{\phi^{E}_{1}} +(1-x_{1})  \overrightarrow{o},\\
 \overrightarrow{\phi^{E}_{3}} &=&(-1-x_{1}) \overrightarrow{\phi^{E}_{1}} +(2+x_{1})  \overrightarrow{o} .
 \end{eqnarray}
Substituting these expressions into the the circular cross product, we obtain the following result:
\begin{equation}
\label{eq:19}
\overrightarrow{\phi^{E}_{1}}\times\overrightarrow{\phi^{E}_{2}}+\overrightarrow{\phi^{E}_{2}}\times\overrightarrow{\phi^{E}_{3}}+\overrightarrow{\phi^{E}_{3}}\times\overrightarrow{\phi^{E}_{1}}
=0\overrightarrow{\phi^{E}_{1}}\times\overrightarrow{o}
=\left(
 \begin{array}{ccc}
   0 & 0 & 0 \\
          \end{array}
            \right)^{T}.
\end{equation}
So the identity also holds in this case.

Employing the identity, we can obtain the area of the EFT. The area vector of the EFT is expressed as
\begin{equation}
\label{eq:20}
\overrightarrow{S}_{\triangle_{EFT}}=\frac{1}{2}(\overrightarrow{\phi^{E}_{2}}-\overrightarrow{\phi^{E}_{1}})\times(\overrightarrow{\phi^{E}_{3}}-\overrightarrow{\phi^{E}_{2}})
=\frac{1}{2}(\overrightarrow{\phi^{E}_{1}}\times\overrightarrow{\phi^{E}_{2}}+\overrightarrow{\phi^{E}_{2}}\times\overrightarrow{\phi^{E}_{3}}+\overrightarrow{\phi^{E}_{3}}\times\overrightarrow{\phi^{E}_{1}})
=\frac{3}{2}Det(\overline{P}^{g})\overrightarrow{o}.
\end{equation}
So the area of the EFT is derived, namely
\begin{equation}
\label{eq:21}
S_{\triangle_{EFT}}=\frac{\sqrt{3}}{2}|Det(\overline{P}^{g})|.
\end{equation}
This formula is called the area theorem in the Ref.\cite{50} in the case that the transition matrix is $\overline{P}$ of the form Eq.\ref{eq:2}.
Our derivations reveal that the theorem results from the the constraints Eq.\ref{eq:9} and  Eq.\ref{eq:10}. Furthermore,
we can see that if two matrices $M_{1}$ and $M_{2}$ meet the constraints, so does their product $M_{1}M_{2}$. Therefore, besides of $\overline{P}$ and $\overline{P}^{g}$, this theorem holds for a general class of flavor transition matrices.
So we call it the generalized area theorem for the EFT. It is useful for the analysis of the uncertainties of the flavor ratio at the source of the HAN.

\subsection{The determinant of $\overline{P}^{g}$ }

The generalized theorem demonstrates that the area of the EFT is determined by the determinant of $\overline{P}^{g}$.
In order to analyse the impacts of the matrix $U^{S}$ on $Det(\overline{P}^{g})$, we show the expression of $Det(\overline{P}^{g})$.  Following the standard parametrization of the leptonic mixing matrix\cite{12}, $U$ is
written as
\begin{equation}
\label{eq:22}
U=
\left(
\begin{array}{ccc}
 c_{12}c_{13} & s_{12}c_{13} & s_{13}e^{-i\delta_{CP}} \\
 -s_{12}c_{23}-c_{12}s_{13}s_{23}e^{i\delta_{CP}} & c_{12}c_{23}-s_{12}s_{13}s_{23}e^{i\delta_{CP}} & c_{13}s_{23} \\
s_{12}s_{23}-c_{12}s_{13}c_{23}e^{i\delta_{CP}} & -c_{12}s_{23}-s_{12}s_{13}c_{23}e^{i\delta_{CP}} & c_{13}c_{23}
\end{array}
\right),
\end{equation}
where $s_{ij}\equiv\sin{\theta_{ij}}$, $c_{ij}\equiv\cos{\theta_{ij}}$.
Here the Majorana phases are not shown because they are irrelevant to the flavor transition matrix.
Similarly, the mixing matrix $U^{S}$ is expressed as
\begin{equation}
\label{eq:23}
U^{S}=
\left(
\begin{array}{ccc}
 c^{S}_{12}c^{S}_{13} & s_{12}^{S}c^{S}_{13} & s^{S}_{13}e^{-i\delta_{CP}^{S}} \\
 -s^{S}_{12}c^{S}_{23}-c^{S}_{12}s^{S}_{13}s^{S}_{23}e^{i\delta^{S}_{CP}} & c_{12}^{S}c_{23}^{S}-s^{S}_{12}s^{S}_{13}s^{S}_{23}e^{i\delta_{CP}^{S}} & c^{S}_{13}s^{S}_{23} \\
s^{S}_{12}s^{S}_{23}-c^{S}_{12}s^{S}_{13}c^{S}_{23}e^{i\delta_{CP}^{S}} & -c^{S}_{12}s^{S}_{23}-s^{S}_{12}s^{S}_{13}c^{S}_{23}e^{i\delta_{CP}^{S}} & c^{S}_{13}c^{S}_{23}
\end{array}
\right),
\end{equation}
where $s^{S}_{ij}\equiv\sin{\theta_{ij}^{}}$, $c^{S}_{ij}\equiv\cos{\theta_{ij}^{S}}$. The superscript S labels  the parameters at the source of the HAN.
According to the expression of $\overline{P}^{g}$, namely Eq.\ref{eq:3}, $Det(\overline{P}^{g})$ is of the form
\begin{equation}
\label{eq:24}
Det(\overline{P}^{g})=Det(U_{sq})Det(U^{S}_{sq}),
\end{equation}
where
\begin{equation}
\label{eq:25}
Det(U_{sq})=(c_{23}^{2}-s_{23}^{2})(c_{12}^{2}-s_{12}^{2})(c_{13}^{2}-s_{13}^{2})+
2s_{13}s_{12}s_{23}c_{12}c_{23}(3s^{2}_{13}-1)\cos\delta_{CP},
\end{equation}
\begin{equation}
\label{eq:26}
Det(U^{s}_{sq})=(c_{23}^{S2}-s_{23}^{S2})(c_{12}^{S2}-s_{12}^{S2})(c_{13}^{S2}-s_{13}^{S2})+
2s^{S}_{13}s^{S}_{12}s^{S}_{23}c^{S}_{12}c^{S}_{23}(3s^{S2}_{13}-1)\cos\delta^{S}_{CP}.
\end{equation}

\subsection{The apparent P violation from  $\overline{P}^{g}$ }

A notable characteristic of the transition matrix  $\overline{P}^{g}$ is the nonsymmetry. The APV could be measured by $\overline{P}^{g}_{\mu e}-\overline{P}^{g}_{e\mu}$, $\overline{P}^{g}_{e\tau}-\overline{P}^{g}_{\tau e}$,
 or $\overline{P}^{g}_{\mu\tau}-\overline{P}^{g}_{\tau \mu}$. According to the unitary conditions, namely Eq.\ref{eq:9} and  Eq.\ref{eq:10}, we find the following identities
\begin{equation}
\label{eq:27}
\overline{P}^{g}_{\mu e}-\overline{P}^{g}_{e\mu}=\overline{P}^{g}_{e\tau}-\overline{P}^{g}_{\tau e}=\overline{P}^{g}_{\tau \mu}-\overline{P}^{g}_{\mu\tau}.
\end{equation}
So we choose $\Delta_{e\mu}\equiv \overline{P}^{g}_{\mu e}-\overline{P}^{g}_{e\mu}$ to describe the APV. In the standard parametrization, $\Delta_{e\mu}$ is  expressed as
\begin{equation}
\label{eq:28}
\Delta_{e\mu}=X\cos\delta_{CP}+Y\cos\delta^{S}_{CP}+Z,
\end{equation}
where $X, Y, Z$ are quantities dependent on the  mixing angles in the matrices $U$ and $U^{S}$.
The relationship between $\Delta_{e\mu}$ and the mixing parameters is obtained with the numerical analysis in the next section.

In order to detect the APV, we should find out the relationship between $\Delta_{e\mu}$ and the flavor ratio vector at Earth.
Note that $\overrightarrow{\phi^{E}}$ is dependent not only on $\overline{P}^{g}$ but also on the specific source of the HAN.
Here we consider the case that the HAN at Earth are from two different sources, namely the pion-decay source $\overrightarrow{\phi^{S_{1}}}$ and the muon-damping source $\overrightarrow{\phi^{S_{2}}}$,
where  $\overrightarrow{\phi^{S_{1}}}$ is of the form
\begin{equation}
\label{eq:29}
\overrightarrow{\phi^{S_{1}}}=\left(
                            \begin{array}{ccc}
                              1/3 & 2/3 & 0\\
                            \end{array}
                          \right)^{T}.
\end{equation}
 $\overrightarrow{\phi^{S_{2}}}$ is just the vector $\overrightarrow{\phi}_{2}$ shown in Eq.\ref{eq:6}. We suppose that the neutrino telescopes in the future could discriminate the sources $S_{1}$
and $S_{2}$ with the help of other astronomical observatories. Furthermore, we assume that the difference between the flavor transition matrix of the source $S_{1}$ and
that of $S_{2}$ is negligible,
namely $\overline{P}^{g}_{S1}$=$\overline{P}^{g}_{S2}$=$\overline{P}^{g}$.
Under this condition, the flavor ratio at Earth with the HAN from $S_{1}$ and  that from $S_{2}$ are expressed respectively as
\begin{equation}
\label{30}
\overrightarrow{\phi^{E_{1}}}=\frac{1}{3}\overrightarrow{\phi^{E}_{1}}+\frac{2}{3}\overrightarrow{\phi^{E}_{2}},~~
\overrightarrow{\phi^{E_{2}}}=\overrightarrow{\phi^{E}_{2}},
\end{equation}
where $\overrightarrow{\phi^{E}_{1}}$ and $\overrightarrow{\phi^{E}_{2}}$ are the column vectors of the matrix $\overline{P}^{g}$ shown in Eq.\ref{eq:8}.
Then the relationship between the APV and the flavor ratios at Earth is obtained, namely
\begin{equation}
\label{eq:31}
\Delta_{e\mu}=3\phi^{E_{1}}_{\mu}-2\phi^{E_{2}}_{\mu}-\phi^{E_{2}}_{e}.
\end{equation}
In general cases, the assumption $\overline{P}^{g}_{S1}$=$\overline{P}^{g}_{S2}$ may not hold . The APV cannot be identified through the measurement of $\overrightarrow{\phi^{E}_{1}}$ and $\overrightarrow{\phi^{E}_{2}}$.
Even so, the term at the right hand of Eq.\ref{eq:31}
could still serve as an index of the $\overline{P}^{g}$ deviation from $\overline{P}$. Namely, if this term is nonzero, we can determine that the flavor transition matrix of the HAN is not$\overline{P}$.

\section{Numerical analyses of impacts of $\overline{P}^{g}$ }
In this section, we perform numerical analyses to study the impacts of the mixing parameters in $\overline{P}^{g}$ on the neutrino flavor ratio at Earth and the APV.
Because of the lack of the information of the source of the HAN, we adopt a model-independent method to describe $U^{S}$.
In order to identify the correlation between the parameters in $U^{S}$ and the new predictions of
$\overline{P}^{g}$, we consider the case that the
relative deviation of $U^{S}$ from the vacuum mixing matrix $U$ is moderate, namely of the order of $10\%$. The specific setup of the parameters of $U^{S}$ is expressed as follows:
\begin{equation}
\label{eq:32}
s^{S2}_{13}=s^{2}_{13}+\Delta s^{2}_{13}, ~~with ~~\Delta s^{2}_{13}\in[-0.02,0.02],
\end{equation}
\begin{equation}
\label{eq:33}
s^{S2}_{12}=s^{2}_{12}+\Delta s^{2}_{12}, ~~with ~~\Delta s^{2}_{12}\in[-0.2,0.2],
\end{equation}
\begin{equation}
\label{eq:34}
s^{S2}_{23}=s^{2}_{23}+\Delta s^{2}_{23}, ~~with ~~\Delta s^{2}_{23}\in[-0.2,0.2],
\end{equation}
\begin{equation}
\label{eq:35}
\delta^{S}_{CP}=\delta_{CP}+\Delta\delta_{CP}, ~~with ~~\Delta \delta_{CP}\in[-1,1]~radian.
\end{equation}
The parameters  $s^{2}_{ij}$ and $\delta_{CP}$ are taken from the global fit data at $3\sigma$ level \cite{4}.
The modifications of the mixing parameters are randomly taken in the given ranges.

\subsection{Uncertainties of the flavor ratios  at the source with the compression map $\overline{P}^{g}$}
We scanned  the determinants of $\overline{P}$ and  $\overline{P}^{g}$ with the mixing parameters given in the above setup.
For the comparison of the impacts of the different parameters, we modify  one parameter at a time.
The correlation of $Det(\overline{P})$ and $Det(\overline{P}^{g})$ is shown in Fig.\ref{fig:2}.
\begin{figure}[t]
\label{fig:2}
\centering 
\includegraphics[width=.49\textwidth]{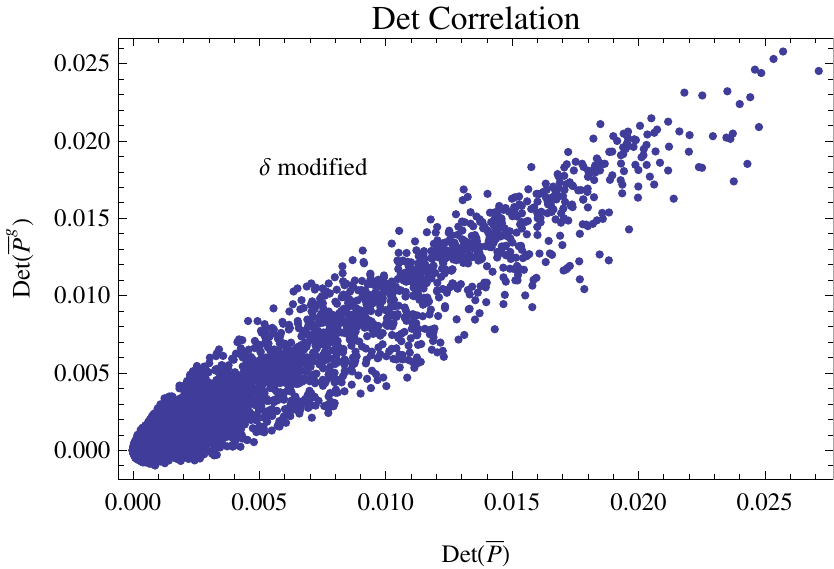}
\hfill
\includegraphics[width=.49\textwidth]{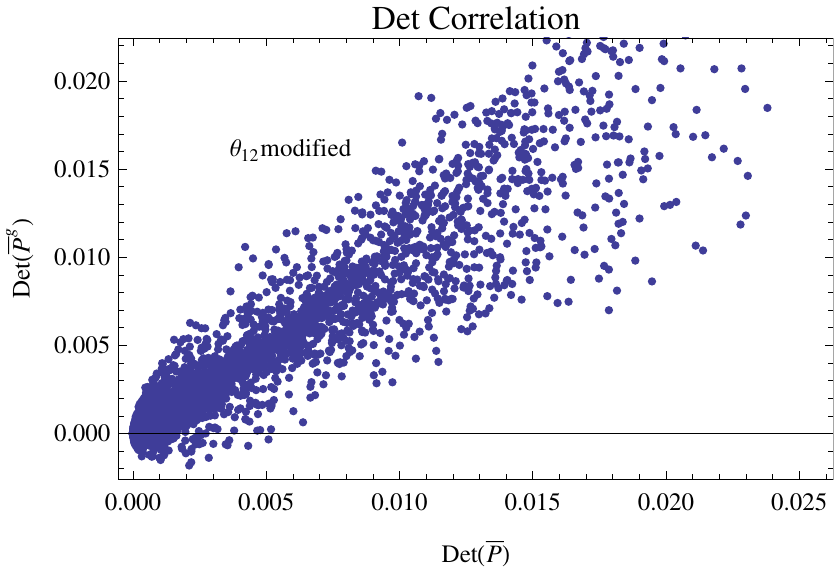}
\hfill
\includegraphics[width=.49\textwidth]{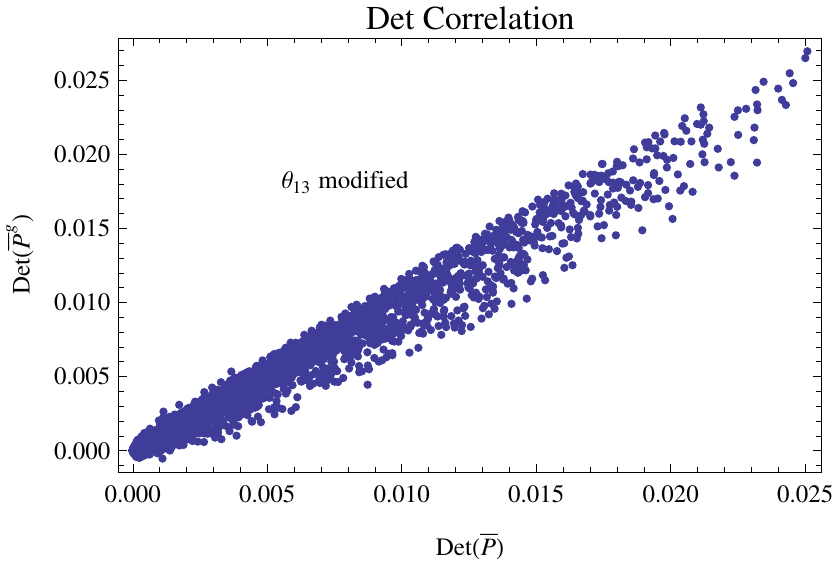}
\hfill
\includegraphics[width=.49\textwidth]{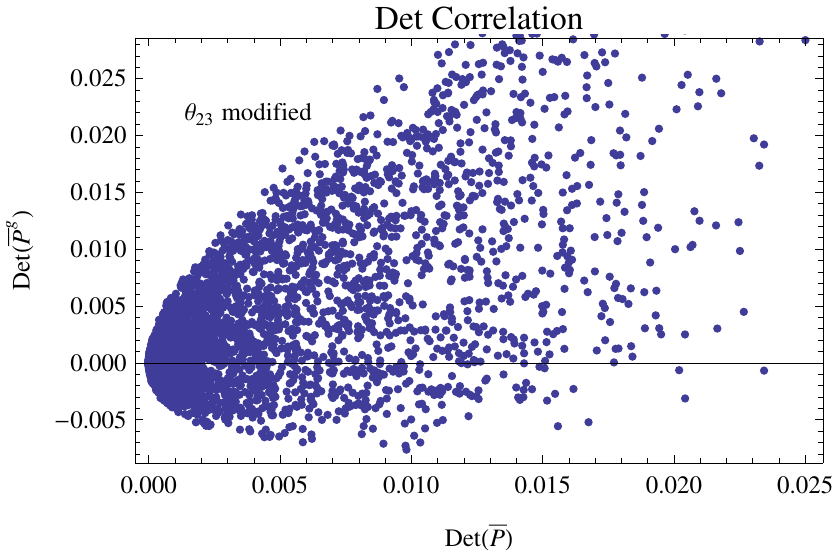}
\caption{\label{fig:2} The correlation between $Det(\overline{P})$ and $Det(\overline{P}^{g})$. The mixing angles and the CP phase in $U$ are in the ranges of the global fit data of the normal mass ordering at $3\sigma$ level \cite{4}.
The results are similar in the case of the inverted mass ordering. The modified parameter in $U^{s}$ is randomly taken in the corresponding range in Eq.\ref{eq:32} - Eq.\ref{eq:35}.  }
\end{figure}
We can see that the range of the magnitude of $Det(\overline{P}^{g})$ is approximate to that of $Det(\overline{P})$.
According to the generalized area theorem for the EFT, the ratio
$S_{\triangle_{EFT}}/S_{\triangle_{FT}}$ is in the range $[0, 0.025]$.
Considering the modification of the mixing parameter in $U^{S}$, the upper bound of the ratio varies in the range $[0.023, 0.028]$.
Therefore, similar to the matrix $\overline{P}$, $\overline{P}^{g}$ is a compression map for the general flavor triangle.
After the flavor transition, the uncertainty of the flavor ratio at the source is decreased  noticeably by  $\overline{P}^{g}$.

For the  quantitative illustration, we decompose the general flavor ratio at the source with the eigenvectors of $\overline{P}^{g}$, i.e.,
\begin{equation}
\label{eq:36}
\overrightarrow{\phi^{S}}=\overrightarrow{o}+x\overrightarrow{v^{1}}+y\overrightarrow{v^{2}},
\end{equation}
where $\overrightarrow{v^{1,2}}$ satisfies the constraint from $\overline{P}^{g}$, namely
$\sum_{i}v^{1,2}_{i}=0$.
The eigen-equations are listed as
\begin{equation}
\label{eq:37}
\overline{P}^{g}\overrightarrow{o}=\lambda^{g}_{0}\overrightarrow{o},~~
\overline{P}^{g}\overrightarrow{v^{1}}=\lambda^{g}_{1}\overrightarrow{v^{1}},
~~\overline{P}^{g}\overrightarrow{v^{2}}=\lambda^{g}_{2}\overrightarrow{v^{2}},
\end{equation}
with the eigenvalues expressed as
\begin{equation}
\label{eq:38}
\lambda^{g}_{0}=1,~~\lambda^{g}_{1,2}=\frac{Tr(\overline{P}^{g})-1}{2}\pm\frac{\sqrt{(Tr(\overline{P}^{g})-1)^{2}-4Det(\overline{P}^{g})}}{2}.
\end{equation}
Here Tr denotes the trace of a matrix. The uncertainty of the flavor ratio at the source could be expressed as
\begin{equation}
\label{eq:39}
\Delta\overrightarrow{\phi^{S}}=\Delta x\overrightarrow{v^{1}}+\Delta y\overrightarrow{v^{2}}.
\end{equation}
After the flavor transition, the corresponding  uncertainty of the flavor ratio at Earth is of the form
\begin{equation}
\label{eq:40}
\Delta\overrightarrow{\phi^{E}}=\lambda^{g}_{1}\Delta x\overrightarrow{v^{1}}+\lambda^{g}_{2}\Delta y\overrightarrow{v^{2}}.
\end{equation}
So the eigenvalues $(\lambda^{g}_{1}, \lambda^{g}_{2})$ can serve as the indices of the compression ratio of the uncertainty of the flavor ratio at the source.
We show the ranges of $(\lambda^{g}_{1},\lambda^{g}_{2})$ in Fig.\ref{fig:3}.
\begin{figure}[t]
\label{fig:3}
\centering 
\includegraphics[width=.49\textwidth]{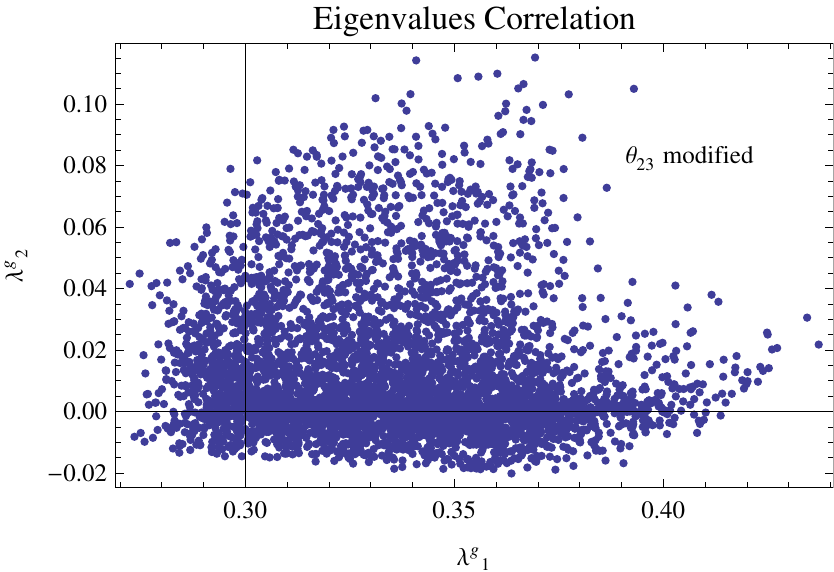}
\hfill
\includegraphics[width=.49\textwidth]{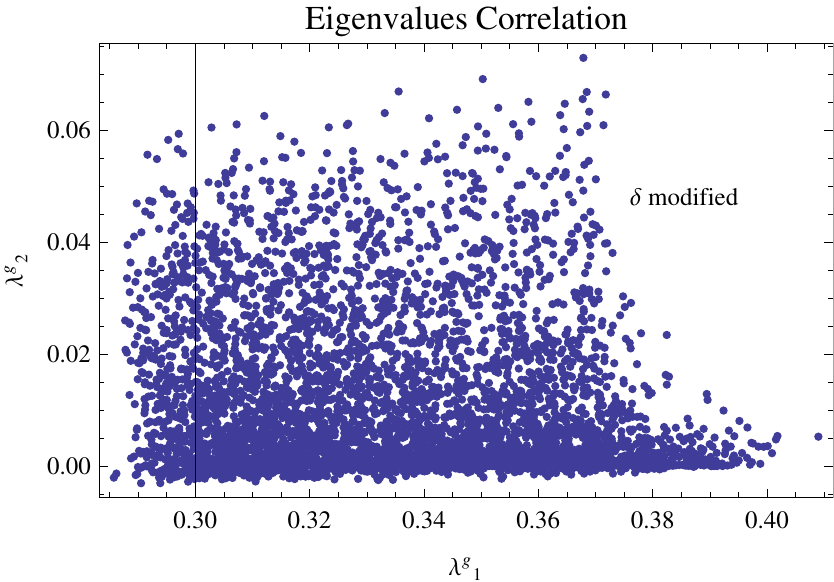}
\caption{\label{fig:3} The eigenvalues of the matrix $\overline{P}^{g}$. The conventions of the mixing parameters are the same as those in Fig.2. }
\end{figure}
We can see that the flavor uncertainty index $\lambda^{g}_{1}\Delta x$ at Earth is around $35\%$ of the initial value.
$\lambda^{g}_{2}\Delta y$ is several percent of the initial value. Particularly, if the eigenvalue $\lambda^{g}_{2}$ is equal to 0, we will lose the
information on the uncertainty of the flavor ratio expressed by  $\Delta y$.
These observations still hold for the matrix $\overline{P}^{g}$ with the modification of $\Delta s^{2}_{12}$ or $\Delta s^{2}_{13}$.
In fact, on the basis of the expression of $Det(\overline{P}^{g})$ shown in Eq.\ref{eq:24} - Eq.\ref{eq:26}, the nature of $\overline{P}^{g}$ as a compression map is determined by the
the ranges of the mixing parameters in $U$. This nature will not be changed with the improvement of the precision of the mixing parameters
 unless $\overline{P}^{g}$ is constructed with the matrix completely different from $U$.

\subsection{Flavor ratio at Earth with the special sources}
Now we analyse the distribution of the flavor ratio at Earth with the special sources.
We consider three typical sources, namely the muon-damping source $\overrightarrow{\phi^{S_{2}}}$=$\left(
                                                                      \begin{array}{ccc}
                                                                        0& 1 & 0 \\
                                                                      \end{array}
                                                                    \right)^{T}$,
the neutron-decay source $\overrightarrow{\phi^{S_{3}}}$=$\left(
                                                                      \begin{array}{ccc}
                                                                        1 & 0 & 0 \\
                                                                      \end{array}
                                                                    \right)^{T}$,
and the pion-decay source  $\overrightarrow{\phi^{S_{1}}}$=$\left(
                                                                      \begin{array}{ccc}
                                                                       1/3& 2/3 & 0 \\
                                                                      \end{array}
                                                                    \right)^{T}$.
The flavor ratios at Earth with these source are shown in Fig.\ref{fig:4} and Fig.\ref{fig:5}.
From these figures, we can obtain following observations:\\
(i)  ~For every source, the impact of $\Delta s_{13}^{2}$ on the distribution of the flavor ratio at Earth is negligible because it is relatively small.
Furthermore, the CP phases $\delta_{CP}$, $\delta^{S}_{CP}$ in the matrix $\overline{P}^{g}$ manifest themselves in the terms $s_{13}\cos\delta_{CP}$ and $s^{S}_{13}\cos\delta^{S}_{CP}$ respectively.
So their impacts on the flavor ratio at Earth are undermined intensively by small 1-3 mixing angles in $U$ and $U^{s}$. The dominant parameters in $\overline{P}^{g}$ are $s_{12}$, $s_{23}$, $s^{S}_{12}$,
$s^{S}_{23}$. We note that a similar observation for the standard matrix $\overline{P}$ has been obtained in the recent Ref.\cite{51}.\\
(ii) ~ For the neutron-decay source and muon-damping source, $s^{S}_{23}$, $s^{S}_{12}$ can enlarge the region of the flavor ratio at Earth noticeably.
However, they cannot exempt the neutron-decay source from the constraint of the $3\sigma$ boundary of the 2015 IceCube data\cite{52}.
 So the new physics in $\overline{P}^{g}$ is still not in favor of the neutron-decay as the unique source of the HAN.  \\
(iii) ~For the pion-decay source, the impacts of $s^{S}_{12}$, $s^{S}_{23}$ on the distribution of the flavor ratio at Earth are weakened by the
approximate $\mu-\tau$ symmetry of the mixing matrix $U$. Here the $\mu-\tau$ symmetry denotes the property $|U_{\mu i}|=|U_{\tau i}|(i=1, 2, 3)$.
As is known, on the basis of the strict $\mu-\tau$ symmetry and the standard transition matrix $\overline{P}$,
the flavor ratio at Earth for the HAN from the the pion-decay source is just  $\overrightarrow{o}=\left(
                                                                                                                                                                     \begin{array}{ccc}
                                                                                                                                                                           1/3 & 1/3 &1/3 \\
                                                                                                                                                                         \end{array}
                                                                                                                                                                       \right)^{T}$\cite{53,54}.
 In our case that $U_{s}$ deviates from $U$ moderately, the flavor ratio at Earth also deviates from the democratic vector  $\overrightarrow{o}$ moderately.\\
 (iv)~~From the light cyan regions in the  figures, we can see that the observations in (i)-(iii) are not changed by the joint impacts of $s^{S}_{12}$, $s^{S}_{23}$ in $\overline{P}^{g}$.

\begin{figure}[t]
\label{fig:4}
\centering 
\includegraphics[width=.49\textwidth]{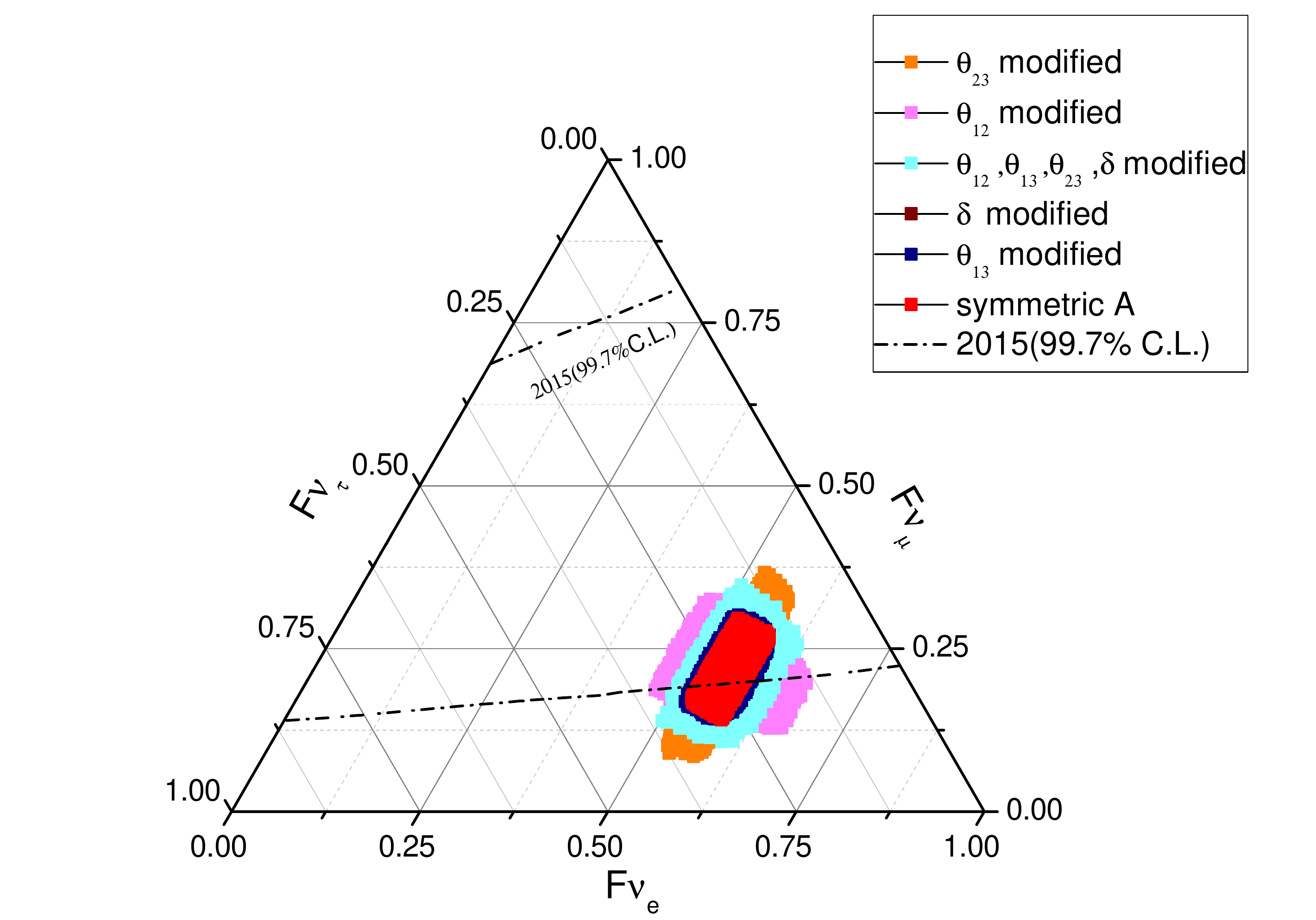}
\hfill
\includegraphics[width=.49\textwidth]{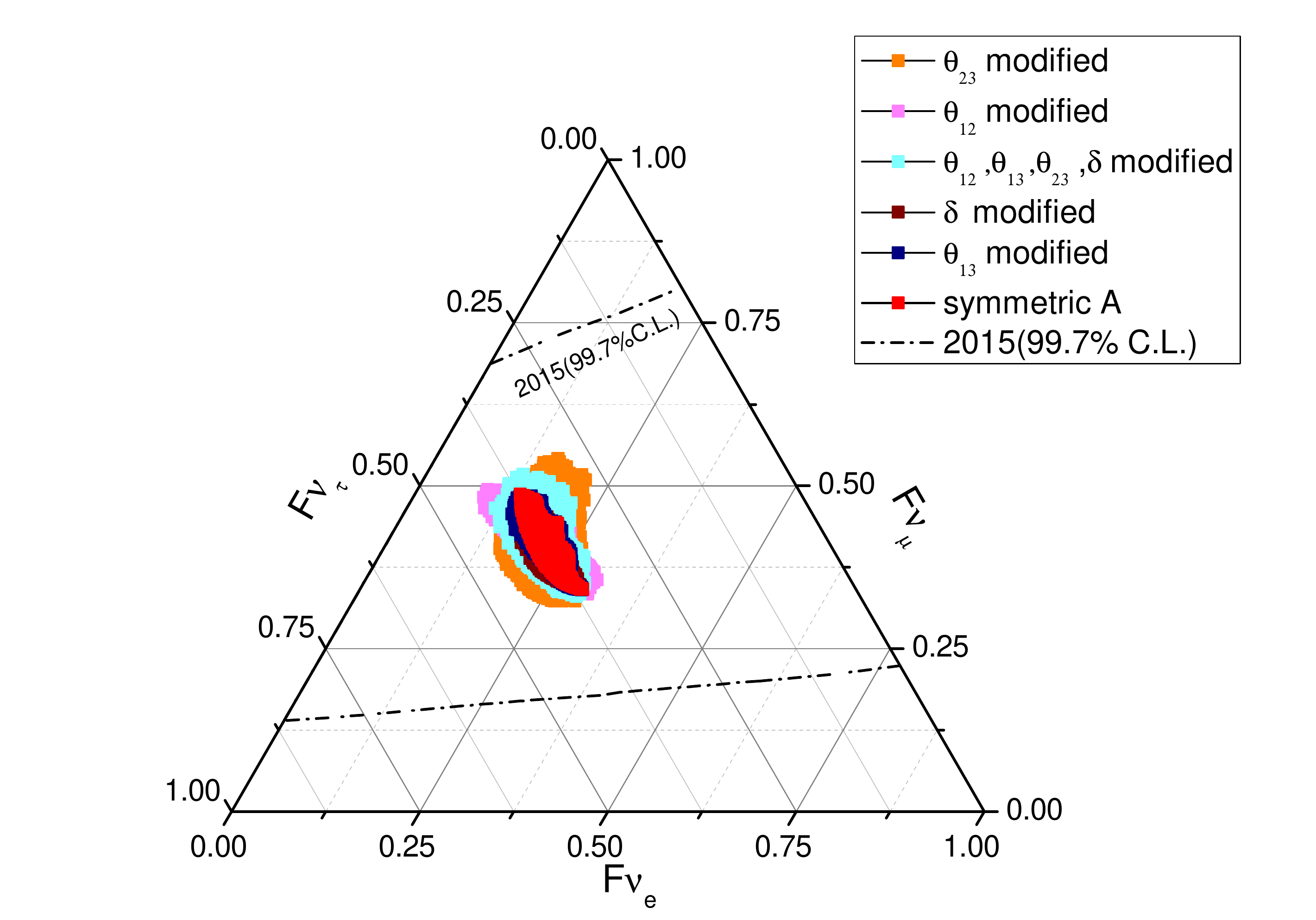}
\caption{\label{fig:4}  The flavor ratio at Earth. The left panel is for the neutron-decay source $(1, 0, 0)$. The right panel is for the muon-damping source $(0, 1, 0)$.
Except for the light cyan region, the conventions of the mixing parameters are the same as those in Fig.2. The light cyan region corresponds to the case that all the mixing parameters are modified in $U^{S}$. The specific
ranges for the modifications are listed as $\Delta s^{2}_{13}\in[-0.01,0.01]$, $\Delta s^{2}_{12}\in[-0.1,0.1]$, $\Delta s^{2}_{23}\in[-0.1,0.1]$, $\Delta \delta_{CP}\in[-0.5,0.5]$ radian. The red region corresponds to
the prediction of the standard matrix $A$. The $3\sigma$ constraint on the flavor ratio is from the 2015 IceCube data\cite{52}.
   }
\end{figure}

\begin{figure}[t]
\label{fig:5}
\centering 
\includegraphics[width=.55\textwidth]{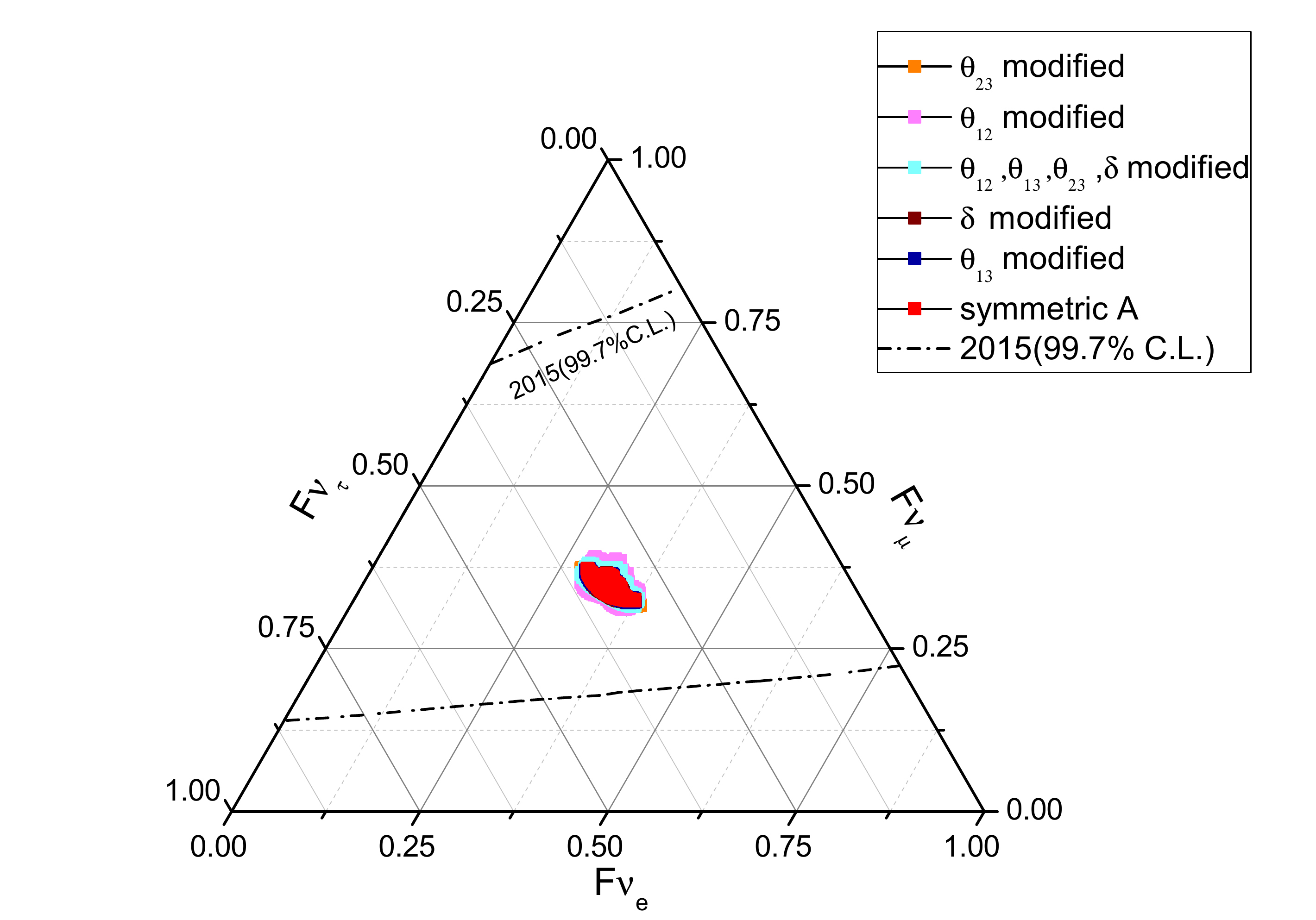}
\caption{\label{fig:5} The flavor ratio at Earth with the pion-decay source $(1/3,~2/3,~0)$. The conventions of the mixing parameters are the same as those in Fig.4. }.
\end{figure}

On the basis of the above observations, our setup of the matrix $\overline{P}^{g}$ shows clear correlations between the distribution of the flavor ratio at Earth
and the parameters of the mixing matrices. The dominant factors in $U^{S}$ are identified, namely $s^{S}_{12}$, $s^{S}_{23}$. Their impacts on the flavor ratios
can be discriminated. So the information on new physics can be read out from the distribution of the flavor ratio at Earth.

\subsection{The apparent P violation modulated by the CP phase}
In principle, the APV results from the mismatch between $U$ and $U^{S}$. The mismatch is quantified by the
modifications $\Delta s_{ij}^{2}$, $\Delta\delta_{CP}$.
The correlations between the APV and the mixing parameters are analysed. We find that the APV
is modulated by the Dirac phase, see Fig.\ref{fig:6} and Fig.\ref{fig:7}.
From the figures, we obtain following observations:\\
(i)~~The APV is modulated by the Dirac phase.
This characteristic results from the expression of  $\Delta_{e\mu}$, see Eq.\ref{eq:28}. In the case $\Delta s_{23}^{2}\neq0$, the oscillation of $\Delta_{e\mu}$ is undermined .
In the case $\Delta s_{23}^{2}=0$, $\Delta\delta_{CP}=0$, we have $\Delta_{e\mu}\sim0$ when $\delta_{CP}\sim\frac{3\pi}{2}$. Correspondingly, $\Delta_{e\mu}$ become maximal when  $\delta_{CP}$ is trivial.
In the case $\Delta\delta_{CP}\neq0$ , $\Delta s_{ij}^{2}=0$, the zero point of  $\Delta_{e\mu}$ is around $\frac{3\pi}{2}+\Delta\delta_{CP}$ or $\frac{\pi}{2}+\Delta\delta_{CP}$.
\\
(ii)~~ The impact of the mass ordering of neutrinos on the APV is not negligible.
 The reason is that the range of $\delta_{CP}$ at $3\sigma$ level with the normal mass ordering is wider than that with the inverted mass ordering, see the specific
 range of $\delta_{CP}$ in Fig.\ref{eq:6} and Fig.\ref{eq:7}. \\
(iii)~~ The oscillating amplitude of $\Delta_{e\mu}$ is determined by the magnitude of $\Delta s_{ij}^{2}$, $\Delta\delta_{CP}$, see Fig.\ref{fig:8} for example.
Compared with $\Delta s_{12}^{2}$, the modification $\Delta s_{23}^{2}$ of the same order of $\Delta s_{12}^{2}$ can lead to the larger amplitude.\\
On the basis of these observation, we find that the magnitude of $\Delta_{e\mu}$ is mainly dominated by  the parameters $\Delta s_{23}^{2}$, $\delta_{CP}$.
Therefore, although $\delta_{CP}$ is not important for the distribution of the flavor ration at Earth, we may still identify its impact on the APV when $\Delta s_{23}^{2}$
is negligible. So in the era of the precise detection the flavor composition of the HAN, we may obtain the information on $\delta_{CP}$ through the secondary effect such as the APV.

\begin{figure}[t]
\label{fig:6}
\centering 
\includegraphics[width=.49\textwidth]{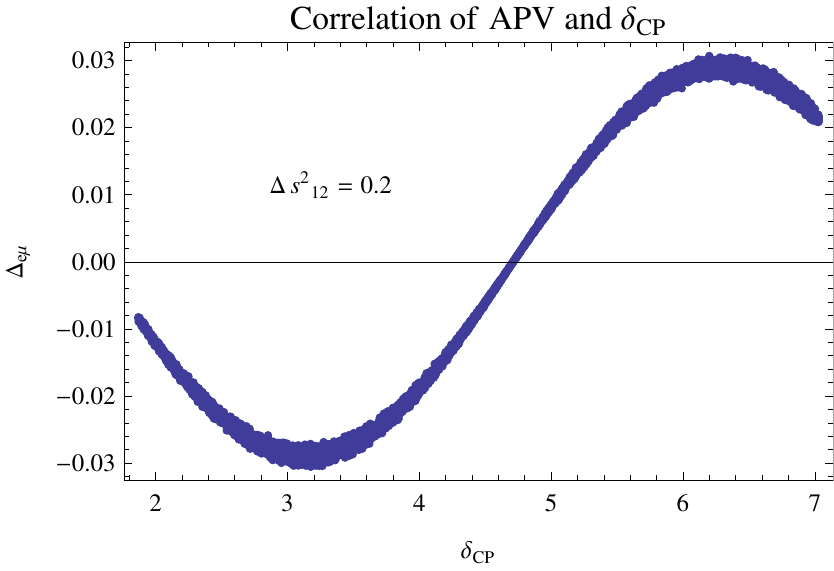}
\hfill
\includegraphics[width=.49\textwidth]{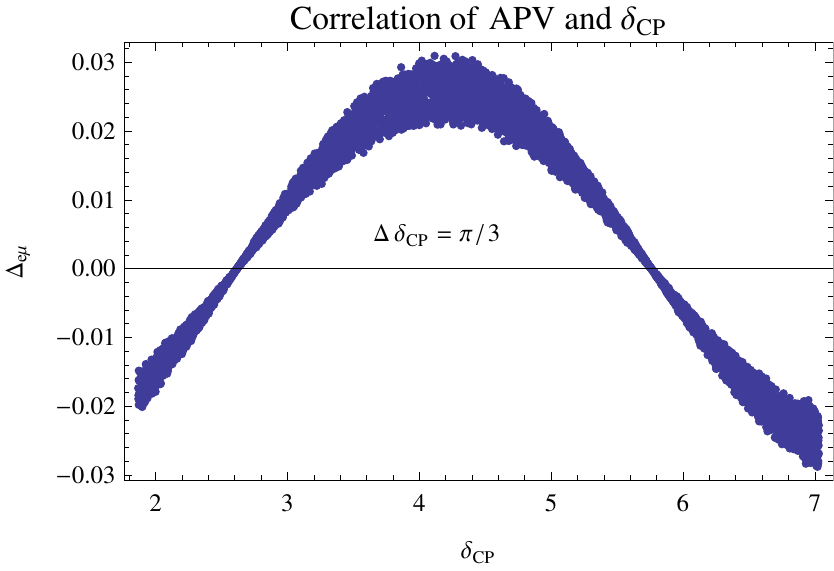}
\hfill
\includegraphics[width=.49\textwidth]{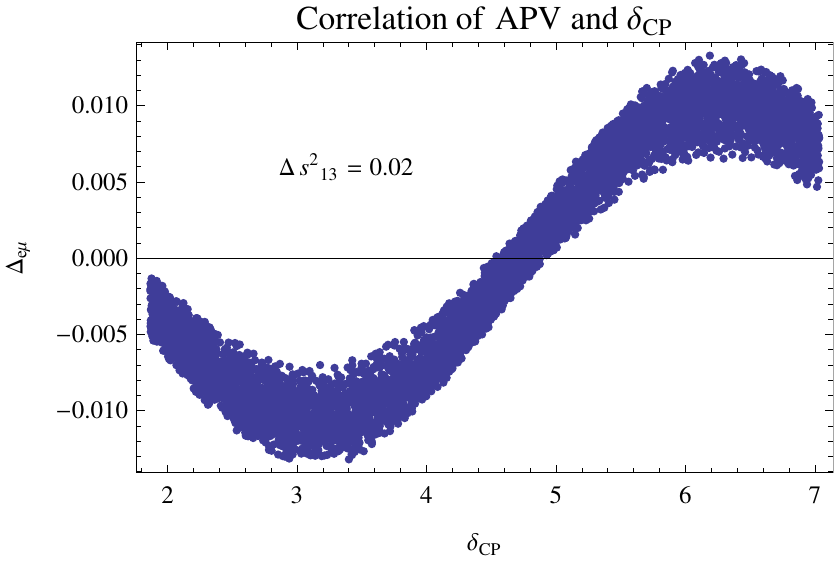}
\hfill
\includegraphics[width=.49\textwidth]{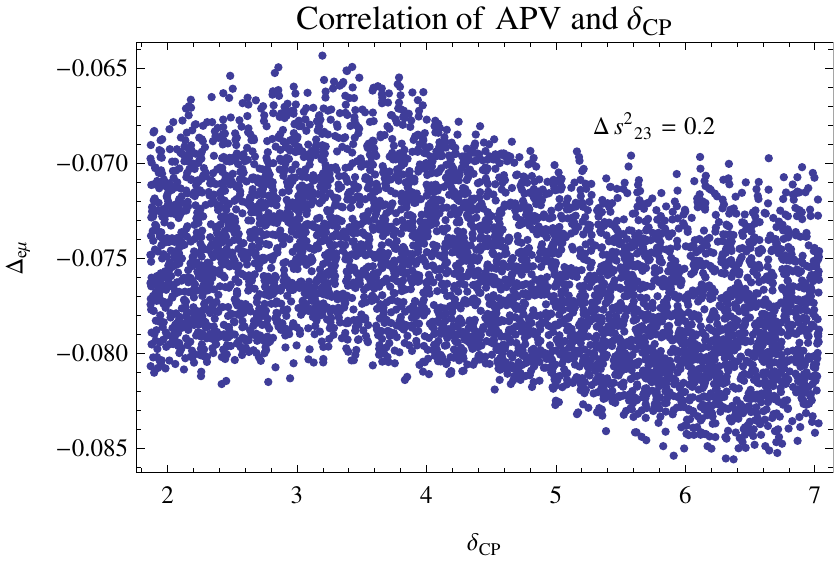}
\caption{\label{fig:6}  The correlation between the apparent P violation(APV) and the CP phase. The mixing parameters in $U$ are in the $3\sigma$ ranges of the global fit data of the normal mass ordering\cite{4}.
 The modified parameter in $U^{s}$ is fixed in every plot. The other parameters in $U^{s}$ are the same as those in $U$.
   }
\end{figure}

\begin{figure}[t]
\label{fig:7}
\centering 
\includegraphics[width=.49\textwidth]{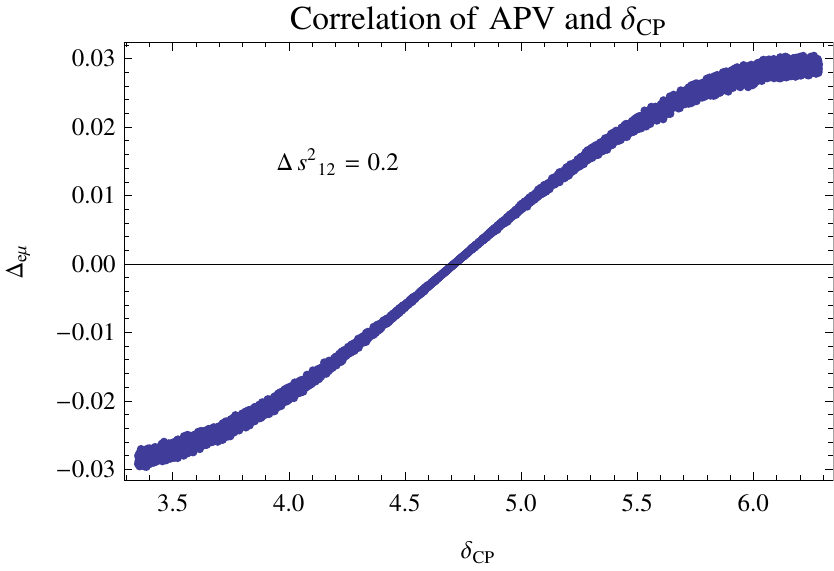}
\hfill
\includegraphics[width=.49\textwidth]{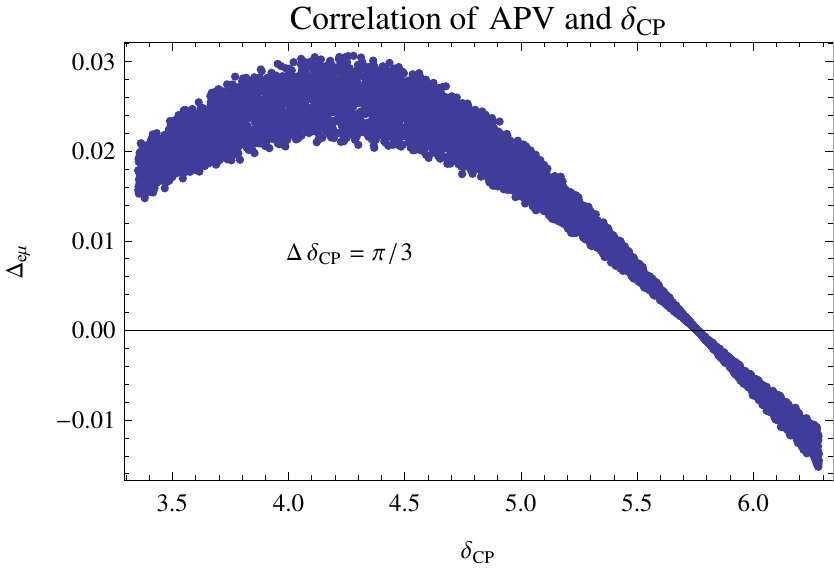}
\hfill
\includegraphics[width=.49\textwidth]{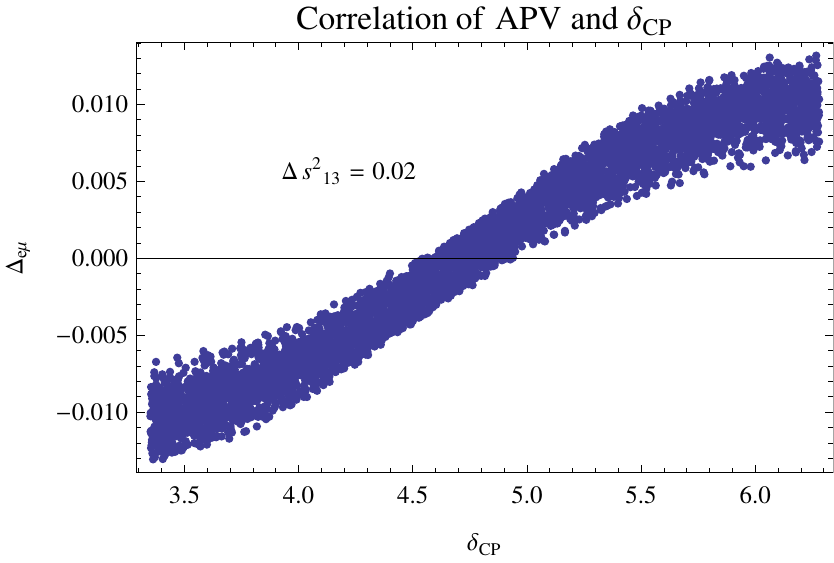}
\hfill
\includegraphics[width=.49\textwidth]{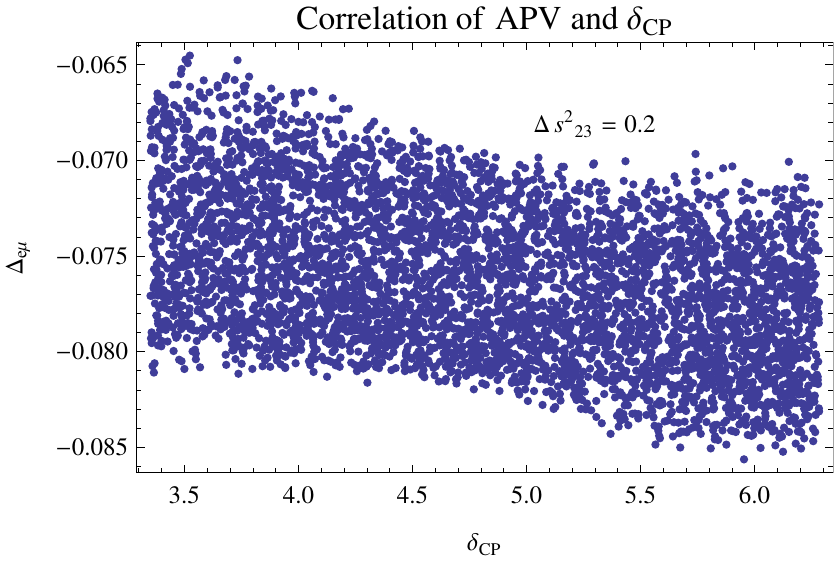}
\caption{\label{fig:7}   The correlation between the apparent P violation(APV) and the CP phase with the inverted mass ordering. The conventions of the modified mixing parameters are the same as those in Fig.6.
   }
\end{figure}

\begin{figure}[t]
\label{fig:8}
\centering 
\includegraphics[width=.49\textwidth]{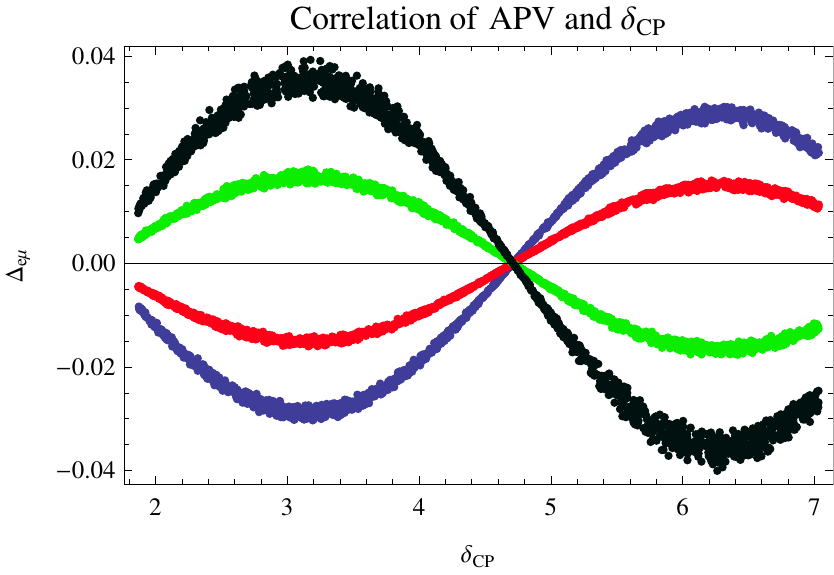}
\hfill
\includegraphics[width=.49\textwidth]{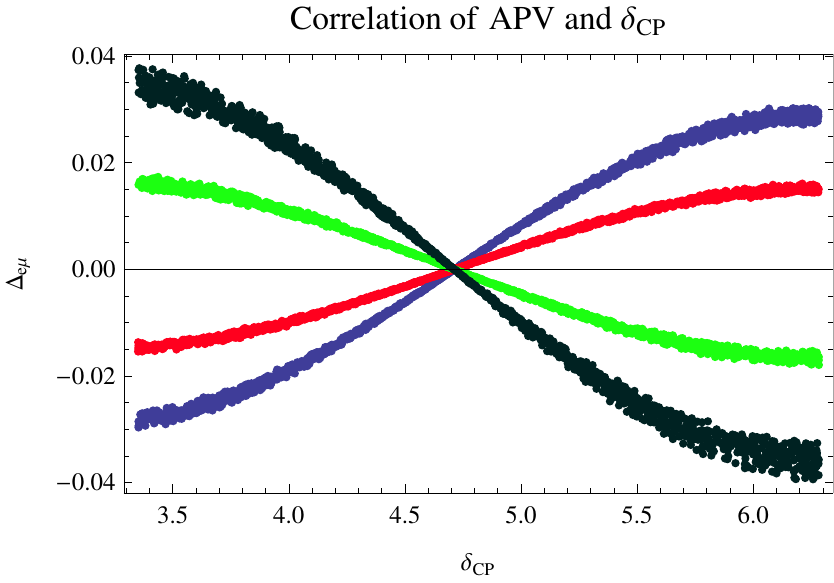}
\caption{\label{fig:8}   The correlation between the apparent P violation(APV) with different amplitudes.
The conventions of the mixing parameters are the same as those in Fig.6.
 The blue scatter points are for the case $\Delta s^{2}_{12}=0.2$, the red points for the case $\Delta s^{2}_{12}=0.1$,
 the green points for the case $\Delta s^{2}_{12}=-0.1$, the black points for the case $\Delta s^{2}_{12}=-0.2$.
 The left panel is for the case of the normal mass ordering.  The right panel is for the case of the inverted mass ordering.
   }
\end{figure}

\subsection{Comment on effects of large modifications in $U^{S}$}

Our observations on the flavor ratios of HAN at Earth and the APV are obtained  in the case that $U^{S}$ differs from $U$ modestly.
When large modification of mixing parameters in $U^{S}$ are considered, we could expect that the region of the flavor ratio distribution of HAN at Earth
for the special source would be expanded noticeably. Furthermore, the correlation between APV and the Dirac CP phase would be disappear under the large modification of
$\Delta s_{23}^{2}$. In special cases that only $\Delta s_{12}^{2}$ and $\Delta s_{13}^{2}$ are relatively large, the correlation could still be identified.

\section{Conclusions}

Inspired by the case of the solar neutrino oscillations, we proposed a generalised matrix $\overline{P}^{g}$ to describe the flavor transition of the HAN.
This matrix is constructed with $U$ in vacuum and $U^{S}$ at the source of the HAN. The mismatch between $U$ and $U^{S}$ leads to the nonsymmetry of $\overline{P}^{g}$
which shows itself through the APV. In order to learn about the impacts of $\overline{P}^{g}$ on the flavor ratio of the HAN at Earth and the APV,
we studied the model-independent properties of $\overline{P}^{g}$ with a vector-analysis method. On the basis of the unitarity of $U$ and $U^{S}$, the generalised area theorem on the EFT
is derived. The area of the EFT is determined by $Det(\overline{P}^{g})$.  With the setup that $U^{S}$ deviates from $U$ moderately,
the impact of $Det(\overline{P}^{g})$ on the uncertainty of the flavor ratio at the source of the
HAN is analysed. The range of $Det(\overline{P}^{g})$ and those of the eigenvalues of $\overline{P}^{g}$ show that the
uncertainty at the source is decreased intensively by $\overline{P}^{g}$ after the flavor transition of the HAN.
The parameters in $U^{S}$ cannot change the nature of $\overline{P}^{g}$ as the compression map of the general flavor triangle.

In the standard parametrization of $U$ and $U^{S}$,
the correlations between the parameters of $U^{S}$ and the distribution of the flavor ratio at Earth
are shown. Compared with the results from the standard flavor transition matrix
$\overline{P}$, we find that the dominant factors in $U^{S}$ for the flavor ratio are $s_{12}^{S}$ and $ s_{23}^{S}$.
Furthermore, the impacts of $ s_{12}^{S}$ and that of $ s_{23}^{S}$ can be discriminated in the ternary plots of the flavor ratios at Earth.
As a secondary effect of the new physics in $U^{S}$, the APV is quantified by $\Delta_{e\mu}$ .
 The correlation between the $\Delta_{e\mu} $ and the mixing parameters in $\overline{P}^{g}$ is obtained. We find that $\Delta_{e\mu}$
is modulated by the Dirac CP phase $\delta_{CP}$. So the information on $\delta_{CP}$ may be obtained from the detection of the APV.
In practice, $\Delta_{e\mu} $ is difficult to detect when the deviation of $U^{S}$ from
$U$ is moderate. We propose a special case that $\Delta_{e\mu} $ may be identified through the flavor ratios at Earth with two distinguishable sources of the HAN,
namely $ \Delta_{e\mu}=3\phi^{E_{1}}_{\mu}-2\phi^{E_{2}}_{\mu}-\phi^{E_{2}}_{e}$. In general cases, this detection scenario may not work. However, the
term $3\phi^{E_{1}}_{\mu}-2\phi^{E_{2}}_{\mu}-\phi^{E_{2}}_{e}$ is still a useful index to test the symmetry of the standard matrix $\overline{P}$ in the future.
Thus, our work may serve as a theoretical preparation for the era of the precise detection of the flavor composition of the HAN.

\vspace{0.08cm}

\acknowledgments
 This work is supported by the National Natural Science Foundation of China under grant No. 12065007, 11705113, the Guangxi Scientific Programm Foundation under grant No. Guike AD19110045, the Research Foundation of Gunlin University of Technology under grant No. GUTQDJJ2018103.\\

\bibliography{refs}

\end{document}